\documentclass[sigconf]{acmart}
\usepackage[utf8]{inputenc}
\renewcommand\footnotetextcopyrightpermission[1]{} 
\usepackage{graphicx}
\usepackage{balance}  
\usepackage{xcolor}

\usepackage{ifthen}
\usepackage{hyperref}
\usepackage{stmaryrd}
\usepackage{amsmath}
\usepackage{colortbl}
\usepackage{balance}
\usepackage{marginnote}
\usepackage{multicol}
\usepackage{listings}
\usepackage[linesnumbered]{algorithm2e}
\usepackage{algpseudocode}
\usepackage{mathtools}
\usepackage{bm}
\usepackage{caption} 
\usepackage{enumerate}
\usepackage{lipsum}
\usepackage{verbatim}

\usepackage{tikz}
\usetikzlibrary{arrows, automata,petri,backgrounds}
\usetikzlibrary{positioning,shapes,shadows,arrows,decorations.markings,snakes}
\usetikzlibrary{decorations.pathmorphing}

\newcommand{\eat}[1]{}

\newcommand\mysection[1]{\vspace{-2mm}\section{#1}\vspace{-0.5mm}}
\newcommand\mysubsection[1]{\vspace{-1mm}\subsection{#1}\vspace{-0.5mm}}
\newcommand\va{\vspace{-0mm}}

\newcommand{\extVersion}{false}  
\newcommand{\printIfExtVersion}[2]
{
        \ifthenelse{\equal{\extVersion}{true}}{#1}{}
        \ifthenelse{\equal{\extVersion}{false}}{#2}{}
}

\copyrightyear{2020}
\acmYear{2020}
\acmConference[BDA 2020]{BDA 2020}{October 2020}{Paris, France}

\begin{document}

\title{Graph integration of structured, semistructured and unstructured data for data journalism}


\author{
Oana Balalau$^1$, Catarina Concei\c{c}\~{a}o$^2$, Helena Galhardas$^2$, Ioana Manolescu$^1$,\\
Tayeb Merabti$^1$,  Jingmao You$^1$, Youssr Youssef$^1$\\
$^1$ Inria, Institut Polytechnique de Paris, \texttt{\small  firstname.lastname@inria.fr}\\
$^2$ INESC-ID and IST, Univ. Lisboa, Portugal,
\texttt{\small{firstname.lastname@tecnico.ulisboa.pt}}
}
\renewcommand{\shortauthors}{O.Balalau, C.~Concei\c{c}ao,
  H.~Galhardas, I. Manolescu, T.~Merabti, J.~You, Y.~Youssef}

\begin{abstract}
Nowadays, journalism is facilitated by the existence of large amounts
of digital data sources, including many Open Data ones. Such data
sources are extremely heterogeneous, ranging from highly structured
(relational databases), semi-structured (JSON, XML, HTML), graphs
(e.g., RDF), and text. Journalists (and other classes of users lacking
advanced IT expertise, such as most non-governmental-organizations, or 
small public administrations) need to
be able to make sense of such heterogeneous corpora, even if they 
lack the ability to define and deploy custom extract-transform-load 
workflows.  These are difficult to set up not only for arbitrary
heterogeneous inputs, but also given that users may want to add (or
remove) datasets to (from) the corpus.

We describe a complete approach for integrating dynamic sets of
heterogeneous data sources along the lines described above:  the
challenges we faced to make such graphs useful, allow their integration to scale, 
and the solutions we proposed for these problems. Our approach
is implemented within the ConnectionLens system; we validate it
through a set of experiments.

\end{abstract}

\maketitle

\mysection{Introduction}
\label{sec:intro}

Data journalists often have to analyze and exploit 
datasets that they obtain from official organizations or their sources, extract from social media, 
or create themselves (typically Excel or Word-style). For instance, journalists from Le Monde newspaper want to retrieve {\em connections between elected people at Assemblée Nationale and companies that have subsidiaries outside of France}. Such a query can be answered currently at a high human effort cost, by inspecting e.g., a JSON list of Assemblée elected officials (available from NosDeputes.fr) and manually connecting the names with those found in a national registry of companies. This considerable effort may still miss connections that could be found if one added information about politicians' and business people's spouses, information sometimes available in public knowledge bases such as DBPedia, or journalists' notes. 

No single query language can be used on such heterogeneous data; instead, we study methods to query the corpus by specifying some keywords and asking for all the connections that exist, in one or across several data sources, between these keywords. 
This line of work has emerged due to our collaboration with Le Monde's fact-checking team\footnote{http://www.lemonde.fr/les-decodeurs/}, within the ContentCheck collaborative research project\footnote{https://team.inria.fr/cedar/contentcheck/}. With respect to the scientific literature, the problem of finding trees that connect nodes matching certain search keywords has been studied under the name of {\em keyword search over structured data}, in particular for relational databases~
\cite{DBLP:journals/debu/YuQC10,DBLP:conf/vldb/HristidisP02}, XML documents~\cite{guo2003xrank,liu2007identifying}, RDF graphs~\cite{DBLP:journals/tkde/LeLKD14,Elbassuoni:2011:KSO:2063576.2063615}.
 However, most of these works assumed one single source of data, in which connections among nodes are clearly identified. When authors considered several data sources~\cite{Li:2008:EEK:1376616.1376706}, they still assumed that one query answer comes from a single data source.

\begin{figure*}[t!]
\centering
\includegraphics[width=.8\textwidth]{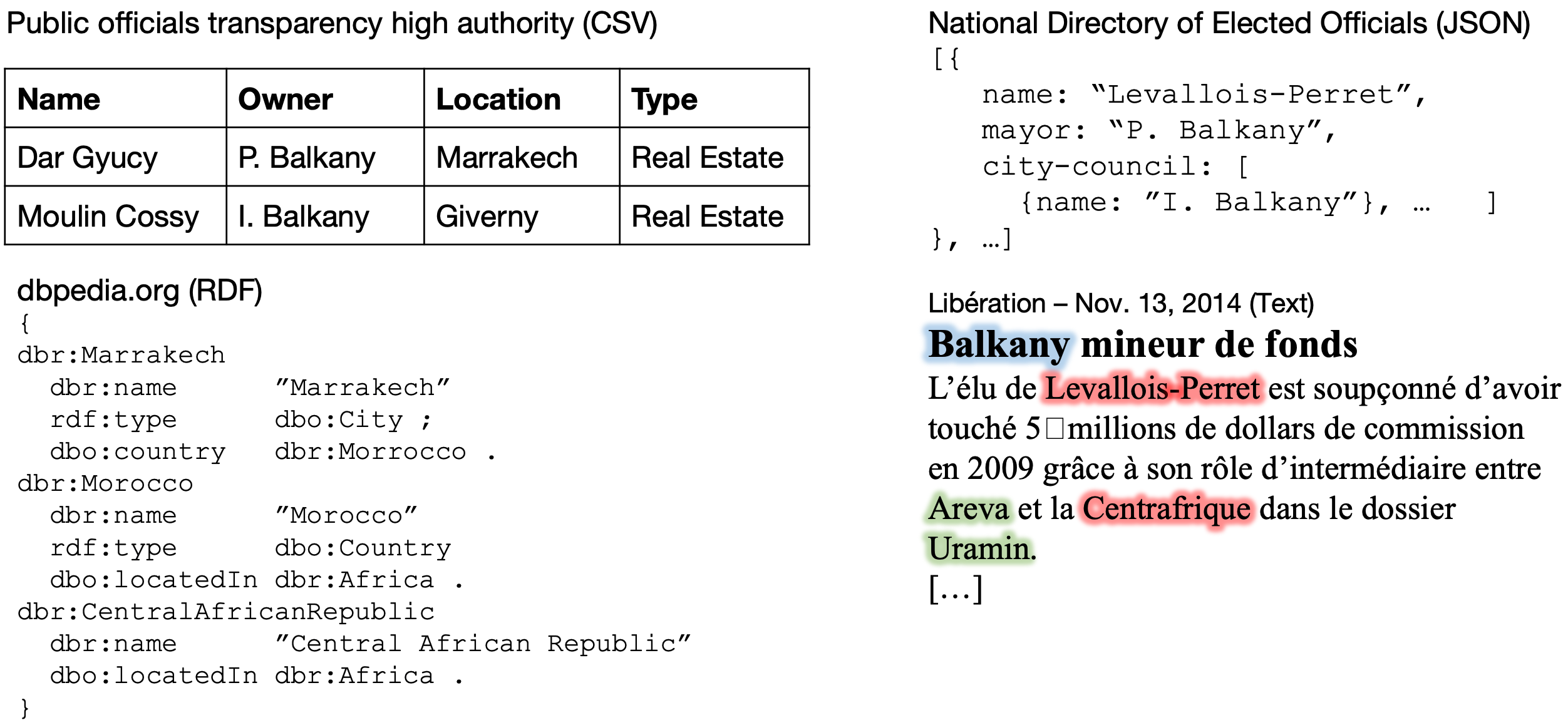}
\vspace{-6.5mm}
\caption{Motivating example: collection ${\mathcal D}$ of four datasets. \label{fig:motiv-example}}
\vspace{-2.5mm}
\end{figure*}

In contrast, the ConnectionLens~\cite{Chanial2018} system (that we are developing since 2018) answers keyword search queries over arbitrary combinations of datasets and heterogeneous data models, independently produced by actors unaware of each other's existence. To achieve this goal, we \textbf{integrate a set of datasets into a unique graph}, subject to the following requirements and constraints: 

\va
\begin{description}
\item[R1. Integral source preservation and provenance:] in journalistic work, it is crucial to be able to trace each node within the integrated graph back to the dataset from which it came. Source preservation is in line with the practice of {\em adequately sourcing} information, an important tenet of quality journalism.
\va
\item[R2. Little to no effort required from users:] journalists often lack time and resources to set up IT tools or data processing pipelines. Even when they may have gained acquaintance with a tool supporting one or two data models (e.g., most relational databases provide some support for JSON data), handling other data models remains challenging.
Thus, the construction of the integrated graph needs to be as automatic as possible.
\va
\item[C1. Little-known entities:] interesting journalistic datasets feature some extremely well-known entities (e.g., highly visible National Assembly deputees such as F.~Ruffin, R.~Ferrand etc.) next to others of much smaller notoriety (e.g., the collaborators employed by the National Assembly to help organize each deputee's work; or a company in which a deputee had worked). From a journalistic perspective, such lesser-known entities may play a crucial role in making interesting connections among nodes in the graph.
\va
\item[C2. Controlled dataset ingestion:] the level of confidence in the data required for journalistic use excludes massive ingestion from uncontrolled data sources, e.g., through large-scale Web crawls.
\va
\item[C3. Language support:]  journalists are first and foremost concerned with the affairs surrounding them (at the local or national scale). This requires supporting dataset in the language(s) relevant for them - in our case, French.
\va
\item[R3. Performance on ``off-the-shelf'' hardware:] Our algorithms' complexity in the data size should be low, and overall performance is a concern;  the tool should run on general-purpose hardware, available to non-expert users like the ones we consider. 
\end{description}
\va

To reach our graph-directed data integration goal under these requirements and constraints, we make the following contributions:

\begin{enumerate}
\va
\item We define the novel {\em integration graphs} we target, and formalize the problem of constructing them from arbitrary sets of datasets.
\va
\item We introduce an  {\em approach}, and an {\em architecture} for building the graphs, leveraging data integration, information extraction,
knowledge bases, and data management techniques. Within this architecture, a significant part of our effort was invested in developing resources and tools for datasets in French. English is also supported, thanks to a (much) wider availability of linguistics resources. 
\va
\item We have fully implemented our approach in an end-to-end tool; it currently supports text, CSV, JSON, XML, RDF, PDF datasets, and existing relational databases.  We present: ($i$)~a set of {\em use cases with real datasets} inspired from our collaboration with journalists; ($ii$)~an {\em experimental evaluation} of its scalability and the quality of the extracted graph.
\end{enumerate}
\va

\noindent\textbf{Motivating example.} To illustrate our approach, we rely on a set of four datasets, shown in Figure~\ref{fig:motiv-example}.  Starting from the top left, in clockwise order, we have: a table with assets of public officials, a JSON  listing of France elected officials, an article from the newspaper Libération with entities highlighted, and a subset of the DBPedia RDF knowledge base. Our goal is to {\em interconnect} these datasets into a graph and to be able to answer, for example, the question: ``What are the connections between Levallois-Perret and Africa?'' One possible answer comes by noticing that P. Balkany was the mayor of Levallois-Perret (as stated in the JSON document), and he owned the ``Dar Gyucy'' villa in Marrakesh (as shown in the relational table), which is in Morocco, which is in Africa (stated by the DBPedia subset). Another interesting connection in this graph is that Levallois-Perret appears in the same sentence
as the Centrafrican Republic in the Libération snippet at the bottom right, which (as stated in DBPedia) is in Africa.

\mysection{Outline and architecture}
\label{sec:outline}

We describe here the main principles of our approach, guided by the
requirements and constraints stated above. 

From requirement R1 (integral source preservation) it follows that
{\em all the structure and content of each dataset  is preserved}
in the integration graph, thus every detail of any dataset is mapped
to some of its nodes and edges. This requirement also leads us to 
{\em preserve the provenance} of each dataset, as well as {\em the
  links that may exist within and across datasets} before loading
them (e.g. interconnected HTML pages, JSON tweets
replying to one another, or RDF graphs referring to a shared resource). 
We term {\em primary nodes} the nodes created in our graph strictly
 based on the input dataset and their provenance; we detail their
 creation in Section~\ref{sec:primary}. 

From requirement R2 (ideally no user input), it follows that  {\em we
  must identify the opportunities automatically to link (interconnect)
  nodes}, even when they were not interconnected in their original
dataset, and even when they come from different datasets. We
achieve this at several levels:

\va
\begin{itemize}
\item We leverage and extend information extraction techniques
  to  {\em extract (identify) entities occurring in the labels of every node in every
    input  dataset}. For instance, ``Levallois-Perret'' is identified
  as a Location in the two datasets at right in
  Figure~\ref{fig:motiv-example} (JSON and Text). Similarly,  ``P. Balkany'', ``Balkany'', ``I. Balkany''
  occurring in the relational, JSON and Text datasets are extracted
  as Person entities. 
 Our method of entity extraction, in particular for the French
 language, is described in Section~\ref{sec:entity-extraction}.
\va
\item We {\em compare (match)} occurrences of entities extracted from the datasets, in order to determine when they refer to the same entity and thus should be interconnected. 
\begin{enumerate}
\item Some entity occurrences we encounter refer to entities known in a {\em trusted
    Knowledge Base} (or KB, in short). For instance, Marrakech,
  Giverny etc. are described in DBPedia; journalists may trust these for such general, non-disputed entities. We  {\em
    disambiguate} each entity occurrence, i.e., try to find the URI
  (identifier) assigned in  the KB to the entity referred to in this occurrence, and we
  {\em connect} the occurrence to the entity. Disambiguation enables, for instance, to
  connect an occurrence of ``Hollande'' to the country, 
  and two other  ``Hollande'' occurrences to the former
    French president. In the latter case, occurrences (thus their datasets) are interconnected.  
We describe the module we built for entity disambiguation
  for the French language (language constraint C3), based on AIDA~\cite{hoffart2011robust}, in
  Section~\ref{sec:disambig}. It is of independent interest, as it can
  be used outside of our context. 
\item On little-known entities (constraint
  C1), disambiguation fails (no URI is found); this is the case, e.g., of ``Moulin
  Cossy'', which is unknown in DBPedia. Combined with constraint C2
  (strict control on ingested sources) it leads to the lack of reliable IDs for many entities mentioned in the datasets. We strive to connect them, as soon as the several identical or at least {\em
    strongly similar} occurrences are found in the same or different datasets. We describe our approach for {\em comparing (matching)} occurrences in
  order to identify identical or similar pairs in
  Section~\ref{sec:matching}. 
\end{enumerate}
\va
\end{itemize}
\va

As the above description shows, our work recalls 
several known areas, most notably data integration, data cleaning, and knowledge base
construction; we detail our positioning concerning these in
Section~\ref{sec:related}. 

\mysection{Primary graph construction from heterogeneous datasets}
\label{sec:primary}


We consider the following \emph{data models}: relational (including SQL databases, CSV files  etc.),  
RDF, 
JSON, XML, or  
HTML, 
and text. 
A dataset $DS=(db, prov)$ in our context is a pair, whose first component is a concrete data object: a relational database, or an RDF graph, or a  JSON, HTML, XML document, or a  CSV, text, or PDF file.  The second component $prov$ denotes the dataset provenance; we consider here that the provenance is a URI, in particular a URL (public or private) from which the dataset was obtained. Users may not wish (or be unable to provide) such a provenance URI when registering a dataset; our approach does not require it, but exploits it when available, as we explain below.

Let $A$ be an alphabet of words.
We define an \textbf{integrated graph} $G=(N,E)$ where $N$ is the set of nodes and $E$ the set of edges. We have $E\subseteq{N\times{N}\times{A^*}\times{[0,1]}}$, where $A^*$ denotes the set of (possibly empty) sequences of words, and the value in $[0,1]$ is the \textit{confidence}, reflecting the probability that the relationship between two nodes holds. 
Each node $n\in N$ has a label $\lambda (n)\in A^*$ and similarly each edge $e$ has $\lambda(e) \in A^*$. We use $\epsilon$ to denote the empty label. We assign to each node and edge a \textbf{unique ID}, as well as a \textbf{type} (simple numbers, unique within one graph $G$). We introduce the supported node types as needed, and write them in bold font (e.g., \textbf{dataset node}, \textbf{URI node}) when they are first mentioned; node types are important as they determine the quality and performance of matching (see Section~\ref{sec:matching}). Finally, we create unique dataset IDs and associate to each node its dataset's ID. 

Let $DS_i=(db_i, prov_i)$ be a dataset of any of the above models. The following two steps are taken regardless of $db_i$'s data model: 
First, we introduce a  \textbf{dataset node} $n_{DS_i}\in N$, which models the dataset itself (not its content). 
Second, if $prov_i$ is not null, we create an \textbf{URI node} $n_{prov_i}\in N$,  whose value is the provenance URI $prov_i$, and an edge $n_{DS_i}\xrightarrow{\text{cl:prov}}n_{prov_i}$, where cl:prov is a special edge label denoting provenance (we do not show these edges in the Figure to avoid clutter).

\begin{figure*}[t!]
\centering
\includegraphics[width=.8\textwidth]{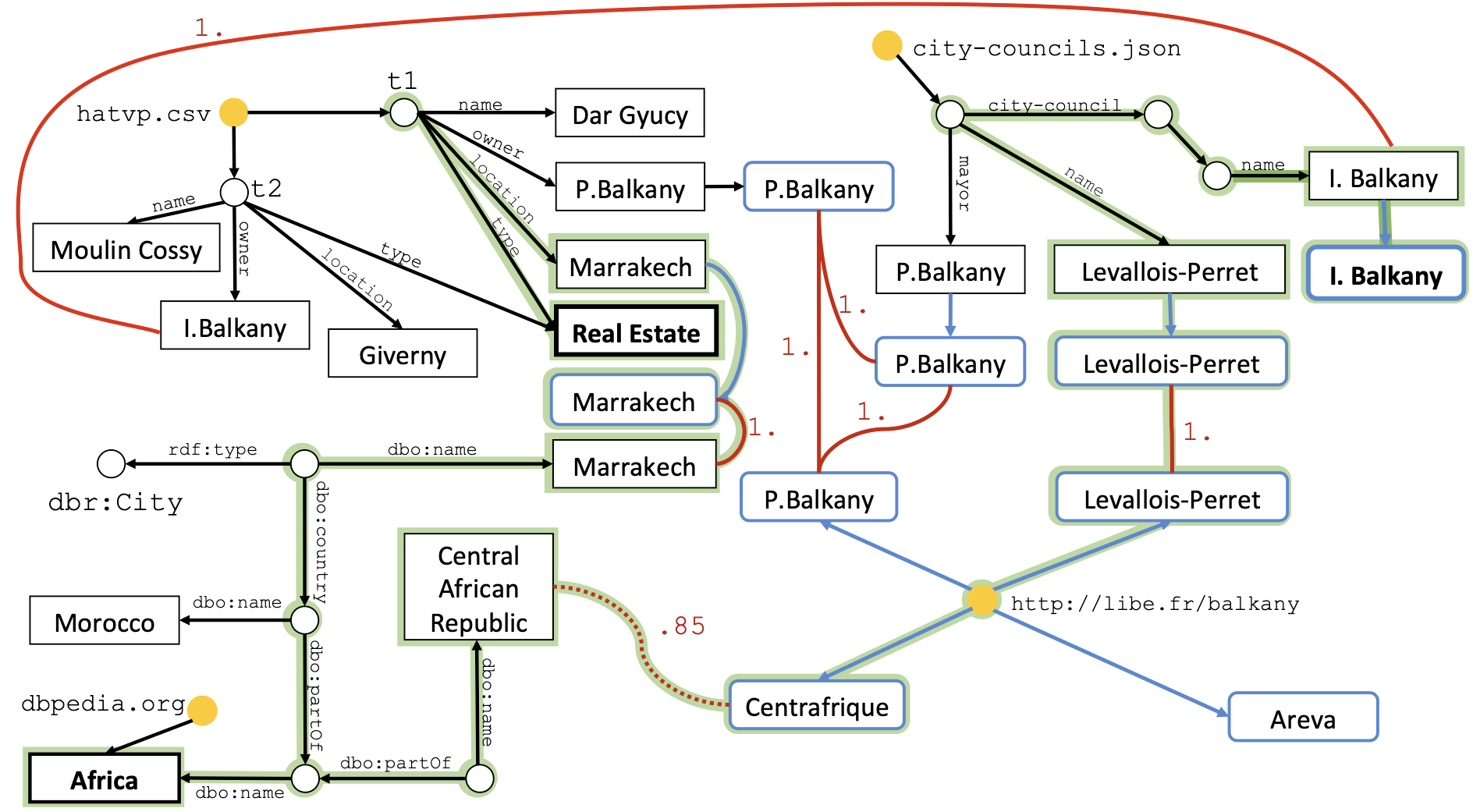}
\vspace{-5mm}
\caption{Integrated graph corresponding to the datasets of Figure~\ref{fig:motiv-example}.  An answer to the keyword query \{``I. Balkany'', Africa, Estate\} is highlighted in light green; the three keyword matches in this answer are shown in bold.\label{fig:example-graph}}
\vspace{-5mm}
\end{figure*}

Next, Section~\ref{sec:datasets-to-graph} explains how each type of dataset yields nodes and edges in $G$.  For illustration, Figure~\ref{fig:example-graph} shows the integrated graph resulting from the datasets in Figure~\ref{fig:motiv-example}. In Section~\ref{sec:graph-refine-optim}, we describe a set of techniques that improve the informativeness and the connectedness and decrease the size of $G$. 
Finally, in Section~\ref{sec:complexity}, we give the complexity of constructing an integrated graph.

\mysubsection{Mapping each dataset to the  graph}
\label{sec:datasets-to-graph}

A common feature of all the edges whose creation is described in this section is that we consider them certain ({\em confidence of $1.0$}), given that they reflect the structure of the input datasets, which we assume trusted. Lower-confidence edges will be due to extraction (Section~\ref{sec:entity-extraction} and Section~\ref{sec:disambig}) and node similarity (Section~\ref{sec:matching}).

\noindent\textbf{Relational.}
\label{sec:vg-relational}
Let $db=R(a_{1},\ldots,a_{m})$ be a relation (table) (residing within an RDBMS, or ingested from a CSV file etc.)
A \textbf{table node} $n_R$  is created to represent $R$ (yellow node with label {\sf hatvp.csv} on top left in Figure~\ref{fig:example-graph}).
Let $t\in R$ be a tuple of the form  $ (v_{1},\ldots,v_{m})$ in $R$. 
A \textbf{tuple node} $n_t$ is created for $t$, with an empty label, e.g., $t_1$ and $t_2$ in Figure~\ref{fig:example-graph}. 
For each non-null attribute $v_i$ in $t$, a \textbf{value node} $n_{v_i}$ is created, together with an edge from $n_t$ to $n_{v_{i}}$, 
labeled $a_{i}$ 
 (for example,  the edge labeled {\sf owner} at the top left in Figure~\ref{fig:example-graph}). To keep the graph readable, confidence values of 1 are not shown.
Moreover, for any two relations $R,R'$ for which we know that attribute $a$ in $R$ is a foreign key referencing $b$ in $R'$, and for any tuples 
 $t \in R$, $t' \in R'$ such that $t.a = t'.b$, the graph comprises an edge from $n_{t}$ to $n_{t'}$ with confidence $1$. This graph modeling of relational data has been used for keyword search in relational databases~\cite{DBLP:conf/vldb/HristidisP02}.

\noindent\textbf{RDF.}
\label{sec:vg-rdf}
The mapping from an RDF graph to our graph is the most natural. 
Each node in the RDF graph becomes, respectively, a URI node or a value node in $G$, and each 
RDF triple $(s, p, o)$ becomes an edge in $E$ whose label is $p$. At the bottom left in Figure~\ref{fig:example-graph} appear some edges resulting from our DBPedia snippet.

\noindent\textbf{Text.} We model a text document very simply, as a node having a sequence of children, where each child is a segment of the text. Currently, each segment is a phrase; we found this a convenient granularity for users inspecting the graph. Any other segmentation could be used instead. 

\noindent\textbf{JSON.}
\label{sec:vg-json}
As customary, we view a JSON document as a tree, with nodes that are either \textbf{map nodes}, \textbf{array nodes} or value nodes. We map each node into a node of our graph and create an edge for each parent-child relation. Map and array nodes have the empty label $\epsilon$. Attribute names within a map become edge labels in our graph. Figure~\ref{fig:example-graph} at the top right shows how a JSON document's nodes and edges are ingested in our graph. 


\noindent\textbf{XML.}
The ingestion of an XML document is very similar to that of JSON
ones. XML nodes are either \textbf{element}, or \textbf{attribute}, or
values nodes. As customary when modeling XML, value nodes are either text children of elements or values of their attributes. 

\noindent\textbf{HTML.} 
\label{sec:vg-html}
An HTML document is treated very similarly to an XML one. In particular, when an HTML document contains a hyperlink of the form \textsf{\small <a href=''http://a.org''>label</a>}, we create a node labeled ``a''  and another labeled ``http://a.org'', and connect them through an edge labeled ``href''; this is the common treatment of element and attribute nodes. However, we detect that a child node satisfies a URI syntax, and {\em recognize (convert) it into a URI node}. This enables us to preserve links across HTML documents ingested together in the same graph,  with the help of node comparisons (see Section~\ref{sec:matching}).

\begin{figure}[h!]
\vspace{-4mm}
  \centering
  \includegraphics[width=.7\columnwidth]{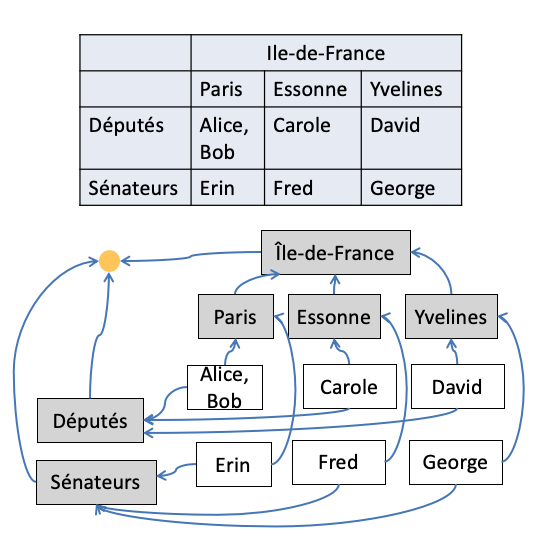}
  \vspace{-4mm}
 \caption{Conversion of a 2d table in a graph.\label{fig:2dtable}} 
\vspace{-3mm}
\end{figure}

\label{sec:bidimensional}
\noindent\textbf{Bidimensional tables.} Tables such as one finds in spreadsheets, which we call bidimensional tables (or 2d tables, in short) differ in important ways from relational database tables. First, unlike a relational table, a 2d table features {\em headers for both the lines and the columns}. 
As spreadsheet users know, this leads to certain flexibility when authoring a spreadsheet, as one feels there is not much difference between a 2d table and its transposed version (where each line becomes a column and vice versa). Second, {\em headers may be nested}, e.g., the 1st column may be labeled ``Paris'', the 2nd column may be labeled ``Essonne'', and a common (merged) cell above them may be labeled ``Ile de France'', as shown in Figure~\ref{fig:2dtable}.

To carry in our graph all the information from a 2d table, and enable meaningful search results, we adopt the approach in~\cite{cao:hal-01583975} for transforming 2d tables into Linked Open Data (RDF). Specifically, we create a \textbf{header cell node} for each header cell, shown as gray boxes at the bottom of Figure~\ref{fig:2dtable}, and a value node for each data cell (white boxes in the figure). Further, each header cell node is connected through an edge to its ancestor header cell, e.g., the edge from the ``Paris'' node to the ``Île-de-France'' node. The edge has a dedicated label cl:parentHeaderCell (not shown in the figure), and each value cell node has edges going to its nearest header cells, edges labeled respectively cl:closestXHeaderCell and cl:closestYHeaderCell. This modeling makes it easy to find, e.g., how Fred is connected to Île-de-France (through Essonne). 
In keeping with~\cite{cao:hal-01583975}, we create a (unique) URI for each data and cell node, attach their value as a property of the nodes, and encode all the edges as RDF triples. 
The approach easily generalizes to higher-dimensional tables. 

\noindent\textbf{PDF.}  A trove of useful data is found in the PDF format. Many interesting PDF documents contain {\em bidimensional tables}. We thus use a dedicated 2d table extractor: the Python Camelot library\footnote{\url{https://camelot-py.readthedocs.io/en/master/}}, amongst the best open source libraries. We then generate from the 2d table a graph representation, as explained above. 
The tables are removed cell by cell from the PDF text, and lines of text are grouped with a rule-based sentence boundary detector so that coherent units are not split. In addition, we add a simple similarity metric to determine the horizontal and vertical headers. Final text content is saved in JSON, parents being the number of the line, children their content. Thus, from a PDF file, we obtain one or more datasets: ($i$)~at least a JSON file with its main text content, and ($ii$)~as many RDF graphs as there were 2d tables recognized within the PDF. From the dataset node of each such dataset $d_i$, we add an edge labeled cl:extractedFromPDF, whose value is the URI of the PDF file. Thus, the PDF-derived datasets are all interconnected, and trace back to their original file. 

\mysubsection{Refinements and optimizations}
\label{sec:graph-refine-optim}

\noindent\textbf{Value node typing.} The need to recognize particular types of values goes beyond identifying URIs in HTML.  URIs also frequently occur in JSON (e.g., tweets), in  CSV datasets etc. 
Thus, we examine each value to see if it follows the syntax of a URI and, if so, convert the value node into a URI one, regardless of the nature of the dataset from which it comes.
More generally, other categories of values can be recognized in order to make our graphs more meaningful. Currently, we similarly recognize \textbf{numeric nodes}, \textbf{date nodes}, \textbf{email address nodes} and \textbf{hashtag nodes}. 

\begin{figure}[th!]
\vspace{-4mm}
\includegraphics[width=7cm]{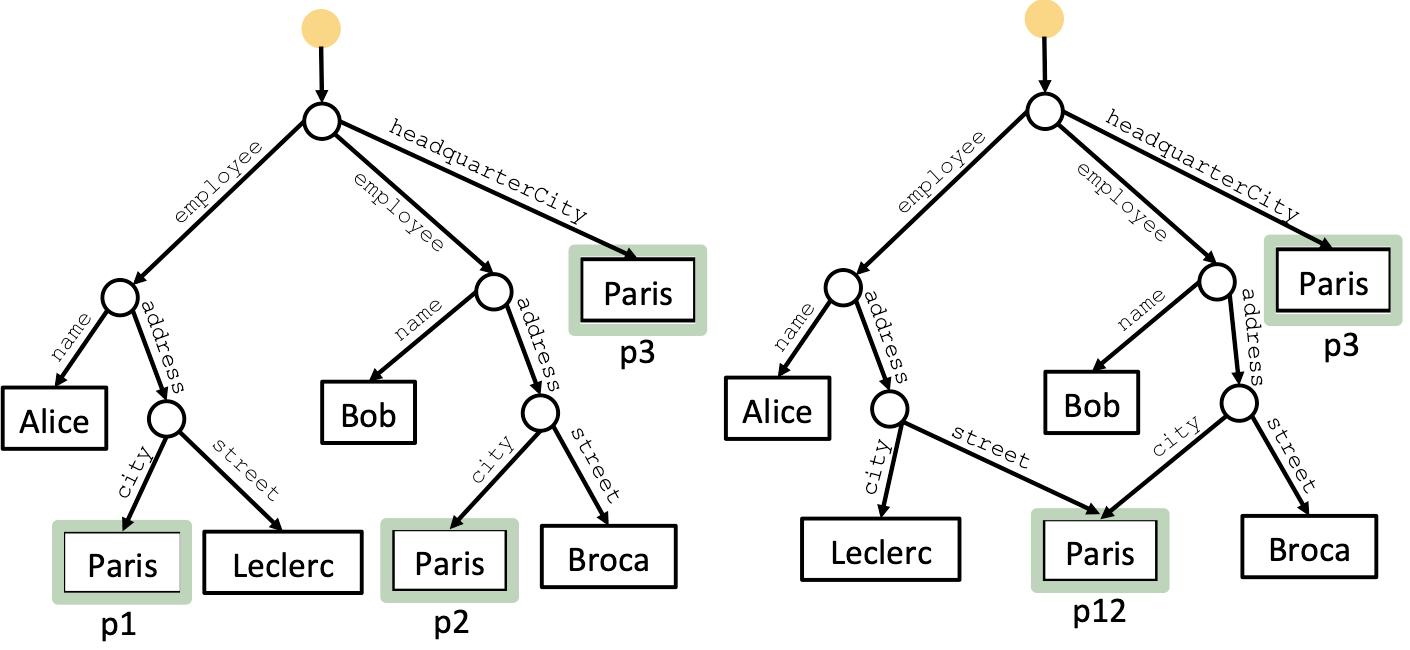}
\vspace{-3.5mm}
\caption{Node factorization example.\label{fig:node-factorization}}
\vspace{-3.5mm}
\end{figure}

\noindent\textbf{Node factorization.} The graph resulting from the ingestion of a JSON, XML, or HTML document, or one relational table, is a tree; any value (leaf) node is reachable by a finite set of label paths from the dataset node. For instance, in the tree at left in Figure~\ref{fig:node-factorization}, two value nodes labeled ``Paris'' (indicated as $p_1,p_2$) are reachable on the paths employee.address.city, while $p_3$ is on the path headquartersCity ($p_3$). Graph creation as described in Section~\ref{sec:datasets-to-graph} creates three value nodes labeled ``Paris''; we call this \textbf{per-occurrence value node creation}. Instead,  \textbf{per-path} creation leads to the graph at right, where a single node $p_{12}$ is created for all occurrences of ``Paris'' on the path employee.address.city. 
We have also experimented with \textbf{per-dataset} value node creation, which in the above example  creates a single ``Paris'' node, and with \textbf{per-graph}, where a single ``Paris'' value node is created in a graph, regardless of how many times ``Paris'' appears across all the datasets. Note that: 

\begin{enumerate}
\item Factorization leads to a DAG (rather than a tree) representation of a tree-structured document; it reduces the number of nodes while preserving the number of edges. 
\item The strongest form of factorization ({\em per-graph})  is consistent with the RDF data model, where a given literal leads to exactly one node. 
\item Factorization leads to more connections between nodes. For instance, at right in Figure~\ref{fig:node-factorization}, Alice and Bob are connected not only through the company where they both work, but also through the city where they live. 
\item Applied blindly on pure structural criteria, factorization may introduce erroneous connections. For instance, constants such as {\em true} and {\em false} appear in many contexts, yet this should not lead to connecting all nodes having an attribute whose value is {\em true}. Another example are named entities, which should be first disambiguated.
\end{enumerate}

To prevent such erroneous connections, we have heuristically identified a set of  \textbf{values which should  not be factorized} even with {\em per-path}, {\em per-dataset} or {\em per-graph} value node creation. Beyond {\em true} and {\em false} and {\em named entities}, this currently includes {\em integer numeric node labels written on less than $4$ digits}, the rationale being that small integers tend to be used for ordinals, while numbers on many digits could denote years, product codes, or other forms of identifiers. This very simple heuristic could be refined. 

Another set of values that should not be factorized are \textbf{null codes}, or strings used to signal missing values, which we encountered in many datasets, e.g., ``N/A'', ``Unknown'' etc. As is well-known from database theory, nulls should not lead to joins (or connections, in our case). We found many different null codes in real-life datasets we considered. We have manually established a small list; we also show to the user \emph{the most frequent values} encountered in a graph (null codes are often among them) and allow her to decide among our list and the frequent values, which to use as null codes. The decision is important because these values will never lead to connections in the graph (they are not factorized, and they are not compared for similarity (see Section~\ref{sec:matching}). This is why we consider that this task requires user input. 


\noindent \textbf{Factorization trade-offs} are as follows. Factorization densifies the connections in the graph. Fewer nodes also reduce the number of comparisons needed to establish similarity links; this is useful since the comparisons cost is in the worst-case quadratic in the number of nodes.  At the same time, factorization may introduce erroneous links, as discussed above. Given that we make no assumption on our datasets, there is no ``perfect'' method to decide when to factorize. The problem bears some connections, but also differences, with entity matching and key finding;  we discuss this in Section~\ref{sec:related}. 

As a pragmatic solution: ($i$)~when loading RDF graphs, in which each URI and each literal corresponds to a distinct node, we apply a {\em per-graph} node creation policy, i.e., the graph contains overall a single node for each URI/literal label found in the input datasets; ($ii$)~in all the other datasets, which exhibit a hierarchical structure (JSON,  XML, HTML, a relational table or CSV file, a text document) we apply the {\em per-dataset} policy, considering that {\em within a dataset}, one constant is typically used to denote only one thing. 

\mysubsection{Complexity of the graph construction}
\label{sec:complexity}

 Summing up the above  processing stages, the time to
ingest a set of datasets in a ConnectionLens graph $G=(N,E)$ is of the
form:

\vspace{-1.5mm}
\begin{center}
$c_1\cdot |E| + c_2\cdot |N| + c_3 \cdot |N_e| +  c_4 \cdot |N|^2$
\end{center}
\vspace{-1.5mm}

In the above, the constant factors are explicitly present (i.e., not
wrapped in an $O(\ldots)$ notation) as the differences between them
are high enough to significantly impact the overall construction time (see Section~\ref{sec:exp-ner} and
\ref{sec:exp-disambiguation}). Specifically: 
 $c_1$ reflects the (negligible) cost of creating each edge using the
 respective data parser, and the (much higher) cost of storing it;
$c_2$  reflects the cost to store a node in the database, and to
invoke the entity extractor on its label, if it is not $\epsilon$;
$N_e$ is the number of entity nodes found in the graph, and $c_3$ is
the cost to disambiguate each entity; finally, the last component
reflects the worst-case complexity of node matching, which may be
quadratic in the number of nodes compared. The constant $c_4$ reflects
the cost of recording on disk that the two nodes are equivalent (one
query and one update) or similar (one update). 

Observe that while $E$ is
entirely determined by the data, the number of value nodes (thus
$N$, thus the last three components of the cist) is impacted by the node creation policy; $N_e$ (and, as we show
below, $c_3$) depend on the entity extractor module used.

\mysection{Named-Entity Recognition}
\label{sec:entity-extraction}

After discussing how to reflect the {\em structure and content} of datasets into a graph, we now look into increasing their {\em meaning}, by leveraging Machine Learning (ML) tools for Information Extraction.

{\em Named entities} (NEs) \cite{nadeau2007ner} are words or phrases which, together, designate certain real-world entities. Named entities include common concepts such as people, organizations, and locations. However, the term can also be applied to more specialized entities such as proteins and genes in the biomedical domain, or dates and quantities. The {\em Named-Entity Recognition} (NER) task consists of $(i)$~identifying  NEs in a natural language text, and  ($ii$)~classifying them according to a pre-defined set of NE types. NER can be modeled as a sequence labeling task, that receives as input a sequence of words and returns a sequence of triples (NE span, i.e., the words referring together to the entity, NE type, $c$) as output, where $c$ is the {\em confidence} of the extraction (a constant between $0$ and $1$). 

Let $n_t$ be a text node. We feed $n_t$ as input to a NER module and create, for each entity occurrence $E$ in $n_t$, an \textbf{entity occurrence node} (or entity node, in short) $n_E$; as explained below, we extract \textbf{Person, Organization} and \textbf{Location} entity nodes. Further, we add an edge from $n_t$ to $n_E$ whose label is cl:extract$T$, where $T$ is the type of $E$, and whose confidence is $c$.  In Figure~\ref{fig:example-graph}, the blue, round-corner rectangles {\sf Centrafrique, Areva, P. Balkany, Levallois-Perret} correspond to the entities recognized from the  text document, while the {\sf Marrakech} entity is extracted from the identical-label value node originating from the CSV file. 

\vspace{3mm}
\noindent\textbf{Named-Entity Recognition} 
We describe here the NER approach we devised for our framework, for English and French.
While we have used Stanford NER~\cite{finkel2005incorporating} in~\cite{Chanial2018,cordeiro:hal-02559688}, we have subsequently developed a more performant module based on the Flair NLP framework~\cite{akbik2019flair}. 
Flair 
 allows {\em training sequence labeling models} using deep-learning techniques. Flair and similar frameworks rely on {\em embedding} words into vectors in a multi-dimensional space. Traditional word embeddings, e.g., Word2Vec \cite{mikolov2013efficient}, Glove \cite{pennington2014glove} and fastText \cite{bojanowski2017enriching}, are \emph{static}, meaning that a word's representation does not depend on the context where it occurs. New embedding techniques are \emph{dynamic}, in the sense that the word's representation also depends on its context. In particular, the Flair dynamic embeddings~\cite{akbik2018contextual} achieve state-of-the-art NER performance. The latest Flair architecture~\cite{akbik2019flair} facilitates {\em combining} different types of word embeddings, as a better performance might be achieved by combining dynamic with static word embeddings. 

The English model currently used in ConnectionLens is a pre-trained model\footnote{https://github.com/flairNLP/flair}. It is trained using the English CoNLL-2003\footnote{https://www.clips.uantwerpen.be/conll2003/ner/} dataset which contains persons, organizations, locations, and miscellaneous (not considered in ConnectionLens) named-entities. The dataset consists of news articles.  The model combines Glove embeddings~\cite{pennington2014glove} and so-called {\em forward and backward pooled} Flair embeddings, that evolve across subsequent extractions. 

To obtain a high-quality model for French, we trained our model on WikiNER~\cite{nothman2013learning},  a multilingual NER dataset automatically created using the text and structure of Wikipedia.
The dataset contains $132$K sentences, $3.4$M tokens and $216$K named-entities, including $74$K Person, $116$K Location and $25$K Organization entities. 
The model uses stacked forward and backward French Flair embeddings with French fastText~\cite{bojanowski2017enriching} embeddings.

\noindent\textbf{Entity node creation.} Similarly to the discussion about value node factorization (Section~\ref{sec:graph-refine-optim}), we have the choice of creating an entity node $n_E$ of type $t$ once per occurrence, or (in hierarchical datasets) {\em per-path}, {\em per-dataset} or {\em per-graph}.   
We adopt the per graph method, with the mention that we will create one entity node for each disambiguated entity and one entity node for each non-disambiguated entity.

\mysection{Entity disambiguation}
\label{sec:disambig}

An entity node $n_e$ extracted from a dataset as an entity of type $T$ may correspond to an entity (resource) described in a trusted knowledge base (KB) such as DBPedia or Yago. As stated in Section~\ref{sec:outline},  this is not always the case, i.e., we encounter entities that are not covered by KBs. However, when they are, the information is valuable as it allows: ($i$)~resolving {\em ambiguity} to make a more confident decision about the entity, e.g., whether the entity node ``Hollande'' refers to the former president or to the country; ($ii$)~tackling {\em name variations}, e.g., two Organization entities labeled ``Paris Saint-Germain Football Club'' and ``PSG'' are linked to the same KB identifier, and ($iii$)~if this is desired, {\em enriching} the dataset with a certain number of facts the KB provides about the entity. 

{\em Named entity disambiguation} (NED, in short, also known as entity linking) is the process of assigning a unique identifier (typically, a URI from a KB) to each named-entity present in a text.  We built our NED module based on  \textbf{AIDA}~\cite{hoffart2011robust}, part of the Ambiverse\footnote{\url{https://www.mpi-inf.mpg.de/departments/databases-and-information-systems/research/ambiverse-nlu/}} framework; AIDA maps entities to resources in YAGO 3~\cite{DBLP:conf/cidr/MahdisoltaniBS15} and Wikidata~\cite{wikidata}.
Our work consisted of ($i$)~adding  support for French (not present in Ambiverse), and ($ii$)~integrating our own NER module (Section~\ref{sec:entity-extraction}) within the Ambiverse framework.


For the first task, in collaboration with the maintainers of Ambiverse\footnote{\url{https://github.com/ambiverse-nlu/ambiverse-nlu\#maintainers-and-contributors}}, we built a new dataset for French, containing the information required for AIDA. The dataset consists of entity URIs, information about entity popularity (derived from the frequency of entity names in link anchor texts within Wikipedia),  and entity context (a set of weighted words or phrases that co-occur with the entity), among others. This information is language-dependent and was computed from the French Wikipedia.

For what concerns the second task,  Ambiverse takes an input text and passes it through a text processing pipeline consisting of \emph{tokenization} (separating words), \emph{part-of-speech (POS) tagging}, which identifies nouns, verbs, etc.,  NER, and finally NED. Text and annotations are stored and processed in Ambiverse using the UIMA standard\footnote{\url{https://uima.apache.org/doc-uima-why.html}}.  A central UIMA concept is the {\em Common Annotation Scheme} (or CAS); in short, it encapsulates the document analyzed, together with all the annotations concerning it, e.g., token offsets, tokens types, etc.  In each Ambiverse module, the CAS object containing the document receives new annotations, which are used by the next module. For example, in the tokenization module, the CAS initially contains only the original text; after tokenization, the CAS also contains token offsets. 

To integrate our Flair-based extractor (Section~\ref{sec:entity-extraction}), we deployed a \textbf{new Ambiverse processing pipeline}, to which we pass as input {\em both the input text and the extracted entities}. 

Due to its reliance on sizeable linguistic resources, the NED module requires $93$G on disk and $32$G of RAM to support disambiguation for English (provided by the authors) and French (which we built). Therefore, we deployed it outside of the ConnectionLens code as a Web service, accessible in our server.

\mysection{Node matching}
\label{sec:matching}

\begin{table*}[t]
\centering
\begin{tabular}{|p{4cm}|p{6.7cm}|c|l|}
\hline
Node group $Gr_1$ &Node pairs from $(Gr_1\times Gr_2)$ to compare &Similarity function&Threshold\\\hline
\hline
Person entity&Person $\times$ Person&Jaro&0.8\\
\hline
Location entity&Location $\times$ Location&Jaro&0.8\\
\hline
Organization entity&Organization $\times$ Organization&Jaro&0.95\\
\hline
URI, hashtag, or email&Same label $\wedge$ same type &Equality&1.0\\
\hline
Number&Same label value $\wedge$ isNumber&Equality&1.0\\
\hline
Date&Same timestamp value $\wedge$ isDate&Equality&1.0\\
\hline
Non-entity short string&abs. len. $<$ 128 $\wedge$ rel. len. $\pm$ 20\% $\wedge$ prefix(3) match&Levenshtein&0.8\\
\hline
Non-entity long string&abs. len. $>$ 32 $\wedge$ rel. len. $\pm$ 20\% $\wedge$ has common word&Jaccard&0.8\\
\hline
\end{tabular}
\caption{Node matching: selectors, similarity functions, and thresholds.\label{fig:comparison-table}}
\vspace{-7mm}
\end{table*}

This section presents our fourth and last method for \textbf{identifying and materializing connections} among nodes from the same or different datasets of the graph. 
Recall that ($i$)~{\em value nodes with identical labels can be fused (factorized)} (Section~\ref{sec:graph-refine-optim});
($ii$)~nodes become connected as {\em parents of a common extracted entity} (Section~\ref{sec:entity-extraction}); 
($iii$)~entity nodes with different labels can be interconnected {\em through a common reference entity} when NED returns the same KB entity (Section~\ref{sec:disambig}).  We still need to be able to connect:

\begin{itemize}
\item \emph{entity nodes with value nodes}, in the case, when extraction did not find anything in the value node. Recall that extraction also uses a context, and it may fail to detect entities on some values where context is lacking. If the labels are identical or very similar, these could refer to the same thing;
\item \emph{entity nodes, such that disambiguation returned no result for one or both}. Disambiguation is also context-dependent; thus, it may work in some cases and not in others; entity nodes with identical or strongly similar labels, from the same or different datasets, could refer to the same thing.
\item \emph{value nodes}, even when no entity was detected. This concerns data-oriented values, such as numbers and dates, as well as text values. The latter allows us to connect, e.g.,  articles or social media posts when their topics (short strings) and/or body (long string) are very similar. 
\end{itemize}

When a comparison finds two nodes with \emph{very similar labels}, we create an edge labeled \textbf{cl:sameAs}, whose confidence is the similarity between the two labels. In Figure~\ref{fig:example-graph}, a dotted red edge (part of the subtree highlighted in green) with confidence
$.85$ connects the ``Central African Republic'' RDF literal node with the
``Centrafrique'' Location entity extracted from the text. 

When a comparison finds two nodes with identical labels, one could consider unifying them, but this raises some modeling issues, e.g., when a value node from a dataset $d_1$ is unified with an entity encountered in another dataset. Instead, we {\em conceptually} connect the nodes with sameAs edges whose confidence is $1.0$. These edges are drawn in solid red lines in Figure~\ref{fig:example-graph}. Nodes  connected by a $1.0$ sameAs edge are also termed {\em equivalent}. Note that $k$ equivalent nodes lead to $k(k-1)/2$ edges. Therefore, these conceptual edges are not stored; instead, the information about $k$ equivalent nodes is stored using $O(k)$ space, as we explain in Section~\ref{sec:storage}.

Inspired by the data cleaning literature, our approach for matching is {\em set-at-a-time}. More precisely, we form node group pairs $(Gr_1, Gr_2)$, and we compare each group pair using the {\em similarity function} known to give the best results (in terms of matching quality) for those groups.  The Jaro measure \cite{jaro1989sim} gives good results for short strings \cite{doan2012int} and is applied to compute the similarity between pairs of entities of the same type recognized by the entity extractor (i.e., person names, locations and organization names which are typically described by short strings). The common edit distance (or Levenshtein distance) \cite{levenshtein1989dist}  is applied to relatively short strings that have not been identified by the entity extractor. Finally, the Jaccard coefficient \cite{jaccard1901sim} gives good results for comparing long strings that have words in common and is therefore used to compute the similarity between long strings.


Table~\ref{fig:comparison-table} describes our matching strategy. For each group $Gr_1$ that we identify, a line in the table shows: a pair $(Gr_1,Gr_2)$ as a logical formula over the node pairs;  the similarity function selected for this case; and the similarity threshold $t$  above which we consider that two nodes are similar enough, and should be connected by a sameAs edge. We set these thresholds experimentally as we found they lead to appropriate similar pair selections, in a corpus built from news articles from Le Monde and Mediapart, social media (tweets), and open (JSON) data from nosdeputes.fr.

For Person, Location, and Organization, the similarity is computed based on normalized versions of their labels, e.g., for Person entities, we distinguish the first name and the last name (when both are present) and compute the normalized label as ``Firstname Lastname''. For URI, hashcode, and email entities, we require identical labels; these can only be found equivalent (i.e., we are not interested in a similarity lower than $1.0$). Finally, we identify {\em short strings} (shorter than 128 characters) and {\em long strings} (of at least 32 characters) and use different similarity functions for each, given that the distinction is fuzzy, we allowed these categories to overlap.

We explain how groups and comparison pairs are formed in Section~\ref{sec:storage} when describing our concrete implementation. 


\mysection{Graph storage}
\label{sec:storage}
The storage module of our platform consists of a Graph interface
providing the data access operations needed in order to access the
graph, and (currently) of a single implementation based on a
relational database, which is accessed through JDBC, and which
implements these operations through a mix of Java code and SQL
queries. This solution has the advantage of relying on
a standard (JDBC) backed by mature, efficient and free tools,
such as Postgres, which journalists were already familiar with, and
which runs on a variety of platforms (requirement R3 from Section~\ref{sec:intro}).  

The table  \textbf{Nodes(\underline{id}, label, type, datasource, label,
normaLabel, representative)} stores the basic attributes of a node,
the ID of its data source, its normalized label, and the ID of its
representative. For nodes not equivalent to any other, the
representative is the ID of the node itself.  As explained previously, the
{\em representative} attribute allows encoding information about
equivalent nodes. 
 Table \textbf{Edges(\underline{id}, source, target, label, datasource, confidence)} stores the
 graph edges derived from the data sources, as well as the extraction
 edges connecting entity nodes with the parent(s) from which they have
 been extracted. Finally, the \textbf{Similar(source, target,
 similarity)} table stores a tuple for each similarity comparison
 whose result is above the threshold (Table~\ref{fig:comparison-table}) but  less than $1.0$. 
The pairs of nodes to be compared for similarity are retrieved by
means of SQL queries, one for each line in the table.  

This implementation requires an SQL query (through JDBC) to access
each node or edge 
within the storage. As such access is expensive, we deploy our own
memory buffer: graph nodes and edges are {\em first created in memory}, and
spilled to disk in batch fashion when the buffer's maximum size is
reached. 
Value node factorization (Section~\ref{sec:graph-refine-optim}) may
still require querying stored graph to check the presence of a
previously encountered label. To speed this up, we also deployed a
memory cache of nodes by their labels for {\em per-graph}
factorization, respectively, by (label, dataset, path) for {\em
per-path} factorization etc. These caches' size determines a compromise
between execution time and memory needs during graph construction.

Other storage back-ends could be 
integrated easily as  implementations of the Graph
interface; we have started experimenting with the BerkeleyDB key-value
store, which has the advantage of running as a Java library, without
the overhead of a many-tiers architecture such as JDBC.

\mysection{Performance evaluation}
\label{sec:experiments}

We now present results of our experiments measuring the performance and the quality of our various modules involved in graph construction.
Section~\ref{sec:exp-settings} presents our  settings, then we present
experiments focusing on: metrics related to graph construction in
Section~\ref{sec:exp-construction}, the quality of our extraction of
tables from PDF documents in Section~\ref{sec:table-extract}, of our
information extraction in Section~\ref{sec:exp-ner}, finally of our
disambiguation module in Section~\ref{sec:exp-disambiguation}.

\mysubsection{Settings}
\label{sec:exp-settings}

The \textbf{ConnectionLens prototype} consists of 46K lines of Java code and 4.6K lines of Python, implementing the module which extracts information from PDF (Section~\ref{sec:primary}) and the  Flair extractor (Section~\ref{sec:entity-extraction}). The latter two have been integrated as (local) Web services. The disambiguator (Section~\ref{sec:disambig}) is Java Web service running on a dedicated Inria server,  adapting the original Ambiverse code to our new pipeline; the code has $842$ classes, 
out of which we modified $10$.  ConnectionLens is available
online at \textbf{\url{https://gitlab.inria.fr/cedar/connectionlens}}. 
We relied on   
 \textbf{Postgres 9.6.5}  to store our integrated graph. 
Experiments ran on a \textbf{server} with  2.7 GHz Intel Core i7 processors and $160$ GB of RAM, running CentOs Linux 7.5.

\vspace{1mm}
\noindent\textbf{Data sources} 
We used a set of diverse real-world datasets, described below from the smaller to the larger. Given that they are of different data models, we order them by their {\em size on disk} before being input in ConnectioneLens.

\noindent\textbf{1.} We built a corpus of 464 \textbf{HTML articles}  crawled from the French online newspaper \href{https://www.mediapart.fr}{Mediapart} with the search keywords ``gilets jaunes'', occupying \textbf{6 MB} on disk.  

\noindent\textbf{2.} An \textbf{XML document}\footnote{https://www.hatvp.fr/livraison/merge/declarations.xml} comprising business interest statements of French public figures, provided by  HATVP ({\em Haute Autorité pour la Transparence de la Vie Publique}); the file occupies \textbf{35 MB}. 

\noindent\textbf{3.} A subset of the \textbf{YAGO 4}~\cite{yago4} RDF knowledge base, comprising entities present in the French Wikipedia and their properties; this takes \textbf{1.21 GB} on disk (9 M triples).  

%
%
%

\mysubsection{Graph construction}
\label{sec:exp-construction}

We start by studying the impact of \textbf{node factorization} (Section~\ref{sec:graph-refine-optim}) on the number of graph nodes and the graph storage time. For that, we rely on the XML dataset, and {\em disable entity extraction, entity disambiguation, and node matching}.
We report the (unchanged) number of edges $|E|$, the number of nodes $|N|$, the time spent storing nodes and edges to disk $T_{DB}$, and the total running time $T$ in Table~\ref{tab:impact}.

\begin{table}[ht!]
\begin{tabular}{@{}p{23mm}rrrrr}
Value node creation policy & $|E|$ & $|N|$ &  $T_{DB}$ (s) & $T$ (s)\\\hline
Per-instance & $1.019.306$ & $1.019.308$ &  $215$ & $225$ \\
Per-path & $1.019.306$ & $514.021$  & $157$& $165$ \\
Per-path w/ null code detection & $1.019.306$ & $630.460$ & $177$ & $187$ \\
Per-dataset & $1.019.306$ & $509.738$ &  $157$ & $167$\\
Per-graph & $1.019.306$ & $509.738$ & $167$ & $177$ \\
Per-graph  w/ null code detection & $1.019.306$ & $626.260$ & $172$ & $181$ \\
\end{tabular}
\caption{Impact of node factorization.}
\label{tab:impact}
\vspace{-5mm}
\end{table}

In this experiment, as expected, the storage time $T_{DB}$  largely dominates the total loading time. 
We also see that $T_{DB}$ is overall correlated with the number of nodes, showing that for this (first) part of graph creation, node factorization can lead to significant performance savings. 
We note that moving from per-instance to per-path node creation strongly reduces the number of nodes. However, this factorization suffers from some errors, as the dataset features many \textbf{null codes} (Section~\ref{sec:graph-refine-optim}); for instance, with per-instance value creation, there are $438.449$ value (leaf) nodes with the empty label $\epsilon$, $47.576$ nodes labeled $0$, $30.396$ ``true'', $26.414$ ``false'', $32.601$ nodes labeled {\em Données non publiées} etc. Using per-path, the latter are reduced to just $95$, which means that in the dataset, the value {\em Données non publiées} appears $32.601$ times on $95$ different paths. However, such factorization (which, as a common child, also creates a connection between the XML nodes which are parents of the same value!) is wrong. When these null codes were input to the tool, such nodes are no longer unified. Therefore the number of nodes increased, also influencing the storage time and, thus, the total time. 
As expected, per-dataset and per-graph produce the same number of nodes (in this case where the whole graph consists of just one dataset), overall the smallest; it also increases when null codes are not factorized. 
We conclude that {\em per-graph value creation combined with null code detection} is a practical alternative.

\begin{figure}
\includegraphics[width=.7\columnwidth]{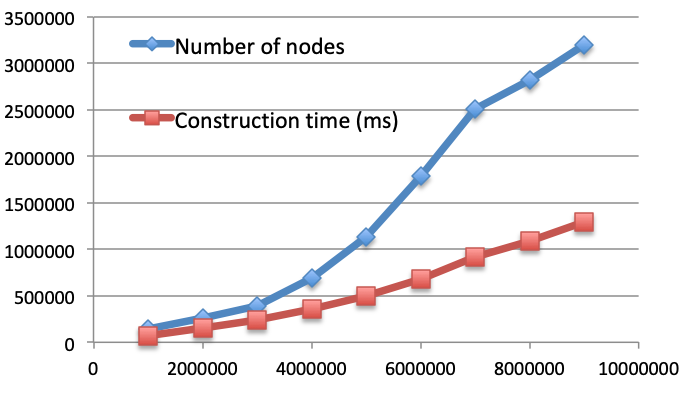}
\vspace{-3mm}
\caption{YAGO loading performance (on the $x$ axis: the number of triples).\label{fig:yago-simple}}
\vspace{-6mm}
\end{figure}

To confirm the \textbf{scalability} of our loading process, we loaded our YAGO subset by slices of 1M triples and recorded after each slice the number of nodes and the running time until that point. Here, we disabled entity extraction, since the graph is already well-structured and well-typed, as well as matching; by definition, each RDF URI or literal is only equivalent to itself. Thus the per-graph value creation policy suffices to copy the RDF graph structure in our graph. Figure~\ref{fig:yago-simple} shows that the loading time grows quite linearly with the data volume.  The figure also shows the cumulated number of different RDF nodes; between 4M and 7M, the number of nodes increases more rapidly, which, in turn, causes a higher cost (higher slope of the loading time) due to more disk writes.

\begin{table*}[t!]
\begin{tabular}{@{}lrrrrrrrrr}
Loading method                       & $|E|$       & $|N|$       &  $T_{DB}$ (s) & $T_E$ (s) & $N_P$    &$N_L$  & $N_O$  &$T_m$ (s) & $T$ (s) \\\hline
No extraction, no matching      & $73.673$ & $61.273$ & $8$              & $0$           & $0$        &$0$       &$0$       &$0$         & $8$ \\
Flair, no matching                    & $90.626$ & $65.686$ & $10$              & $2.019$    &$1.535$  &$1.240$ &$1.050$ &$0$        &$2.029$ \\
Flair (with cache), full matching   &$90.626$  & $65.686$ &$10$             & $1.923$    &$1.535$  &$1.240$ &$1.050$ &$2.976$ &$3.178$ \\
Flair (with cache), entity-only matching&$90.626$  & $65.686$ &$10$             & $1.923$    &$1.535$  &$1.240$ &$1.050$ &$2.021$ &$2.224$ \\
\end{tabular}
\caption{Loading the Mediapart HTML graph (entity extraction results were cached in the last two rows).\label{tab:html}}
\vspace{-5mm}
\end{table*}

Next, we quantify the impact of \textbf{extraction} (Section~\ref{sec:entity-extraction}) and of \textbf{matching} (Section~\ref{sec:matching}). For this, we rely on the HTML corpus, which is rich in mentions of people, locations and organizations. For reference, we first load this graph without extraction, nor matching. Then, we load it with Flair extraction, and, respectively: without matching; with matching as described in Section~\ref{sec:matching}; and with matching from which we exclude the node pairs where no node is an entity (that is, the ``short string'' and the ``long string'' comparisons).  Table~\ref{tab:html} shows the results. Besides the metrics already introducesd, we report: $T_E$, the time to extract entities; the numbers $N_P$, $N_L$, and $N_O$ of extracted people, locations and organization entities; and the matching time $T_m$. The latter includes the time to update the database to modify node representatives (this also requires two queries) and to create similarity edges. Table~\ref{tab:html} shows the first Flair entity extraction dwarves the data storage time. Second, the corpus features a significant number of entities (more than $4.000$). Third, matching is also quite expensive, due to the computation of the similarity functions on pairs of node labels and the many read/write operations required in our architecture. Finally, a constant aspect in all the experiments we performed is that {\em the largest share of similarity comparison costs is incurred by comparisons among (long and short) strings}, not entities; this is because strings are more numerous. Non-entity strings rarely match, and the most significant connections among datasets are on entities. Therefore, we believe good-quality graphs can be obtained even without string comparisons. In Table~\ref{tab:html}, avoiding it allows us to significantly reduce the total time by about $33\%$.

\mysubsection{Table extraction from PDF documents}
\label{sec:table-extract}

Our approach for identifying and extracting bidimensional tables from PDF documents (Section~\ref{sec:bidimensional}) customizes and improves upon the Camelot library to better detect data cells and their headers. We present here an evaluation of the quality of our extraction, using metrics adapted from the literature~\cite{gobel2012methodology}. Specifically,  based on a corpus of PDFs labeled with ground truth values~\cite{gobel2013icdar}, we measure the precision and recall of three tasks: ($i$) \textit{table recognition} (REC) is measured on the list of tokens recognized as belonging to the table; ($ii$)~\textit{table structure detection} (STR) is measured on the adjacency relations of cells, i.e. each cell is represented by its content, and its nearest right and downwards neighbors; ($iii$)~\textit{table interpretation} (INT) quantifies how well are represented the headers of a cell; a cell is correctly {\em interpreted} if its associated header cells are correctly identified. Precision and recall are measured to reflect the full or partial matching of headers. 

\begin{table}[h!]
\vspace{-2mm}
\begin{tabular}{|l|c|c|c|}\cline{2-4}
\multicolumn{1}{l|}{pdfXtr}
        & Precision &  Recall  & $F1$ \\\hline
LOC     &  70.16\% & 76.06\% &  72.02\% \\
STR     &   71.48\% &  73.52\% &  71.68\% \\
INT    &   40.51\% &  40.01\% &  39.64\% \\\hline
Overall &   60.72\% &  63.20\% &  61.12\% \\\hline
\end{tabular}
\caption{Quality of pdfXtr extraction algorithm.\label{fig:pdf-results-table}}
\vspace{-8mm}
\end{table}

We aggregate the measures over the $55$ documents contained in the above-mentioned dataset and we show the results in Table~\ref{fig:pdf-results-table}.
The table shows that the quality of table recognition and table structure is quite good, though not allowing yet for full automation. The relatively low performance on the last task can be explained by the variety of styles in header construction, which could be better tackled by a supervised learning algorithm.

\mysubsection{Named-Entity Recognition}
\label{sec:exp-ner}

Due to the unavailability of an off-the-shelf, good-quality entity extractor for French text, we decided to train a new model.
To decide the best NLP framework to use, we experimented with the Flair~\cite{akbik2019flair} and SpaCy (\url{https://spacy.io/})  frameworks. Flair allows {\em combining} several embeddings, which can lead to significant quality gains. Following~\cite{akbik2019flair}, after testing different word embedding configurations and datasets, we trained a Flair model using {\em stacked forward and backward French Flair embeddings} with {\em French fastText embeddings} on the WikiNER dataset. We will refer to this model as \textit{Flair-SFTF}.


Below, we describe a \emph{qualitative comparison} of \textit{Flair-SFTF}  with the French Flair and SpaCy {\em pre-trained} models.
The French pre-trained Flair model 
is trained with the WikiNER dataset, and uses French character embeddings trained on Wikipedia, and French fastText embeddings.
As for SpaCy, two pre-trained models are available for French: a medium (\textit{SpaCy-md}) and a small one (\textit{SpaCy-sm}). They are both trained with the WikiNER dataset and the same parameterization. The difference is that \textit{SpaCy-sm} does not include word vectors, thus, in general,  \textit{SpaCy-md}  is expected to perform better, since word vectors will most likely impact positively the model performance. 
Our evaluation also includes the model previously present in ConnectionLens~\cite{Chanial2018}, trained using \textit{Stanford NER}~\cite{finkel2005incorporating}, with the Quaero Old Press Extended Named Entity corpus~\cite{galibert2012extended}. 

We measured the precision, recall, and $F1$-score of each model using the \textit{conlleval} evaluation script, previously used for such tasks\footnote{The script \url{https://www.clips.uantwerpen.be/conll2002/ner/} has been developed and shared in conjunction with the CoNLL (Conference on Natural Language Learning).}. \textit{conlleval} evaluates \emph{exact matches}, i.e.,  both the text segment of the proposed entity and its type, need to match ``gold standard'' annotation, 
 to be considered correct. Precision, recall, and $F1$-score (harmonic mean of precision and recall) are computed for each named-entity type.
To get an aggregated, single quality measure, \textit{conlleval} computes the {\em micro-average} precision, recall, and $F1$-score over all recognized entity instances, of all named-entity types. 

For evaluation, we used the entire FTBNER dataset~\cite{sagot-etal-2012-annotation}.
We pre-processed it to convert its entities from the seven types they used, to the three we consider, namely, persons, locations and organizations. 
After pre-processing, the dataset contains $12$K sentences 
and $11$K named-entities ($2$K persons, $3$K locations and $5$K organizations).

\begin{table}[h!]
\begin{tabular}{|l|c|c|c|}\cline{2-4}
\multicolumn{1}{l|}{Flair-SFTF}
        & Precision &  Recall  & $F1$ \\\hline
LOC     &   59.52\% &  79.36\% &  68.02\% \\
ORG     &   76.56\% &  74.55\% &  75.54\% \\
PER     &   72.29\% &  84.94\% &  78.10\% \\\hline
Micro &   69.20\% &  77.94\% &  73.31\% \\\hline
\multicolumn{4}{c}{}\\\cline{2-4}
\multicolumn{1}{l|}{Flair-pre-trained}
        & Precision &  Recall  & $F1$ \\\hline
LOC     &   53.26\% &  77.71\% &  63.20\% \\
ORG     &   74.57\% &  75.61\% &  75.09\% \\
PER     &   71.76\% &  84.89\% &  77.78\% \\\hline
Micro &   65.55\% &  77.92\% &  71.20\% \\\hline
\multicolumn{4}{c}{}\\\cline{2-4}
\multicolumn{1}{l|}{SpaCy-md}
        & Precision &  Recall  & $F1$ \\\hline
LOC     &   55.77\% &  78.00\% &  65.04\% \\
ORG     &   72.72\% &  54.85\% &  62.53\% \\
PER     &   53.09\% &  74.98\% &  62.16\% \\\hline
Micro &   61.06\% &  65.93\% &  63.40\% \\\hline
\multicolumn{4}{c}{}\\\cline{2-4}
\multicolumn{1}{l|}{SpaCy-sm}
        & Precision &  Recall  & $F1$ \\\hline
LOC     &   54.92\% &  79.41\% &  64.93\% \\
ORG     &   71.92\% &  53.23\% &  61.18\% \\
PER     &   57.32\% &  79.19\% &  66.50\% \\\hline
Micro &   61.25\% &  66.32\% &  63.68\% \\\hline
\multicolumn{4}{c}{}\\\cline{2-4}
\multicolumn{1}{l|}{Stanford NER}
        & Precision &  Recall  & $F1$ \\\hline
LOC     &   62.17\% &  69.05\% &  65.43\% \\
ORG     &   15.82\% &   5.39\% &   8.04\% \\
PER     &   55.31\% &  88.26\% &  68.00\% \\\hline
Micro &   50.12\% &  40.69\% &  44.91\% \\\hline
\end{tabular}
\caption{Quality of  NER from French text.\label{fig:ner-results-table}}
\vspace{-8mm}
\end{table}

The evaluation results are shown in Table \ref{fig:ner-results-table}.
All models perform better overall than the \textit{Stanford NER} model previously used in ConnectionLens~\cite{Chanial2018,cordeiro:hal-02559688}, which has a micro $F1$-score of about 45\%. 
The \textit{SpaCy-sm} model has a slightly better overall performance than \textit{SpaCy-md}, with a small micro $F1$-score difference of $0.28\%$. \textit{SpaCy-md} shows higher $F1$-scores for locations and organizations, but is worse on people, driving down its overall quality. 
All Flair models surpass the micro scores of SpaCy models. In particular, for people and organizations, Flair models show more than $11\%$ higher $F1$-scores than SpaCy models. Flair models score better on all named-entity types, except for locations when comparing the SpaCy models, specifically, with the \textit{Flair-pre-trained}.
\textit{Flair-SFTF} has an overall $F1$-score of $73.31\%$ and has better scores than the \textit{Flair-pre-trained} for all metrics and named-entity types, with the exception of the recall of organizations, lower by $1.06\%$.
In conclusion, {\em Flair-SFTF} is the best NER model we evaluated. 

Finally, we study {\em extraction speed}. The average time to extract named-entities from a sentence is: for Flair-SFTF $22ms$, Flair-pre-trained $23ms$, SpaCy-md $9ms$, SpaCy-sm  $9ms$, and Stanford NER $1ms$. 
The quality of Flair models come at a cost: they take, on average, more time to extract named-entities from sentences.
SpaCy extraction is about twice as fast, and Stanford NER much faster. Note that {\em extraction time is high, compared with other processing costs};  as a point of comparison, on the same hardware,  {\em tuple access on disk through JDBC} takes  $0.2$ to $1$ ms (and in-memory processing is of course much faster). This is why \textbf{extraction cost is often a significant component} of the total graph construction time.

\mysubsection{Disambiguation}
\label{sec:exp-disambiguation}
We now move to the evaluation of the disambiguation module. 
As mentioned in Section~\ref{sec:disambig}, our module works for both English and French. 
The performance for English has been measured on the CoNLL-YAGO dataset~\cite{hoffart2011robust}, by the developers of Ambiverse. They report a micro-accuracy of $84.61\%$ and a macro-accuracy of $82.67\%$.
To the best of our knowledge, there is no labeled corpus for entity disambiguation in French, thus we evaluate the performance of the module on the FTBNER dataset previously introduced.
FTBNER consists of sentences annotated with named entities. The disambiguation module takes a sentence, the type, and offsets of the entities extracted from it, and returns for each entity either the URI of the matched entity or an empty result if the entity was not found in the KB.  In our experiment,  $19\%$ of entities have not been disambiguated, more precisely $22\%$ of organizations, $29\%$ of persons, and $2\%$ of locations. 
For a fine-grained error analysis, we sampled 150 sentences and we manually verified the disambiguation results (Table~\ref{fig:ned}). The module performs very well, with excellent results for locations ($F1 = 98.01\%$), followed by good results for organizations ($F1 = 82.90\%$) and for persons ($F1 = 76.62\%$). In addition to these results, we obtain a micro-accuracy of $90.62\%$ and a macro-accuracy of $90.92\%$. The performance is comparable with the one reported by the Ambiverse authors for English. We should note though that the improvement for French might be due to our smaller test set. 

\begin{table}[h!]
\vspace{-2mm}
\begin{tabular}{|l|c|c|c|}\cline{2-4}
\multicolumn{1}{l|}{}
        & Precision &  Recall  & $F1$\\\hline
LOC     &   99.00\% &  97.05\% &  98.01\% \\
ORG     &   92.38\% &  75.19\% &  82.90\% \\
PER     &   75.36\% &  77.94\% &  76.62\% \\\hline
Micro &   90.51\% &  82.94\% &  86.55\% \\\hline
\end{tabular}
\caption{Quality of disambiguation for French text.\label{fig:ned}}
\vspace{-2mm}
\end{table}

The average time of disambiguation per entity is $347$ ms, which is \textbf{very significant} (again when compared with database access through JDBC of $.2$ to $2$ ms). This is mostly due to the fact that the disambiguator is deployed as a Web service, incurring a client-server communication overhead for making a WSDL call. At the same time,  as explained in Section~\ref{sec:disambig}, the disambiguation task itself involves large and costly resources; its cost is high, and directly correlated with the number of entities in the datasets. In our current platform, users can turn it on or off.

\noindent\textbf{Experiment conclusion} Our experimens have shown that
heterogeneous datasets can be integrated with robust performance in persistent graphs, to
be explored and queried by users. We will continue work to optimize
the platform. 
\mysection{Related work and perspectives}
\label{sec:related}

Our work belongs to the area of {\em data integration}~\cite{doan2012int}, whose goal is to facilitate working with different databases (or datasets) as if there was only one. Data integration can be achieved either in a {\em warehouse} fashion (consolidating all the data sources into a single repository), or in a {\em mediator} fashion (preserving the data in their original repository, and sending queries to a mediator module which distributes the work to each data source, and finally combines the results). Our initial proposal to the journalists we collaborated with was a mediator~\cite{DBLP:journals/pvldb/BonaqueCCGLMMRT16}, where users could write queries in an ad-hoc language, specifying which operations to be done at each source (using, respectively, SQL for a relational source, SOLR's search syntax for JSON, and SPARQL for an RDF source). This kind of integration through a source-aware language has been successfully explored in polystore systems~\cite{kolev2016cloudmdsql,alotaibi:hal-02070827}. In the past, we had also experimented with a mixed XML-RDF language for fact-checking applications~\cite{goasdoue:hal-00814285,goasdoue:hal-00828906}. However, the journalists' feedback has been that the installation and maintenance of a mediator over several data sources, and querying through a mixed language, were very far from their technical skills.  This is why here, we ($i$)~pursue a {\bf warehouse} approach; ($ii$)~base our architecture on {\bf Postgres}, a highly popular and robust system; ($iii$)~simplify the query paradigm to keyword querying.

Because of the applicative needs, ConnectionLens integrates a wide variety of data sources: JSON, relational, RDF and text since~\cite{Chanial2018}, to which, since our report~\cite{cordeiro:hal-02559688}, we added XML, multidimensional tables, and the ability to extract information from PDF documents. We integrate such heterogeneous content in a  graph; from this viewpoint, our work recalls the production of {\em Linked Data}. A significant difference is that {\em we do not impose that our graph is RDF}, and {\em we do not assume, require, or use a domain ontology}. The latter is because adding data to our tool should be as easy as dropping a JSON, XML or text file in an application; the overhead of designing a suitable ontology is much higher and requires domain knowledge. 
Further, journalists with some technical background found ``generic'' graphs, such as those stored in Neo4J, more straightforward and more intuitive than RDF ones; this discourages both converting non-RDF data sources into RDF and relying on an ontology. 
Of course, our graphs could be exported in RDF;  and, in different application contexts, 
Ontology-Based Data Access~\cite{DBLP:conf/cikm/Lenzerini11,DBLP:journals/pvldb/CalvaneseGLV12,buron:hal-02446427} brings many benefits. Another difference is the pervasive application of  Information Extraction on all the values in our graph; we discuss Information Extraction below. 

Graphs are also produced when {\em constructing knowledge bases}, e.g., Yago~\cite{DBLP:conf/cidr/MahdisoltaniBS15,yago4}. Our settings is more limited in that we are only allowed to integrate a given set of datasets that journalists trust. Harvesting information from the World Wide Web or other sources whose authorship is not well-controlled was found risky by journalists who feared a ``pollution'' of the database. Our choice has therefore been to {\em use\/} a KB only for disambiguation and accept (as stated in Section~\ref{sec:intro}) that the KB does not cover some entities found in our input datasets. Our simple techniques for matching (thus, connecting) nodes are reminiscent of data cleaning, entity resolution~\cite{DBLP:journals/pvldb/SuchanekAS11,DBLP:conf/edbt/0001IP20}, and key finding in knowledge bases, e.g.~\cite{DBLP:journals/kbs/SymeonidouAP20}. Much more elaborate techniques exist, notably, when the data is regular, its structure is known and fixed, an ontology is available, etc.; none of these holds in our setting. 

{\em Pay-as-you-go data integration} refers to techniques whereby data is initially stored with little integration effort and can be better integrated through subsequent use. This also recalls 
 "database cracking"~\cite{Idreos:2007,Alagiannis:2012}, where the storage is re-organized to better adapt to its use. 
We plan to study adaptive stores as part of our future work. 

 {\em Dataspaces\/} were introduced in~\cite{Franklin:2005} to designate ``a large number of diverse, interrelated data sources''; 
The authors note the need to support data sources of heterogeneous data models, and blend query and search, in particular through support for keyword search. 
Google's Goods system~\cite{Halevy:2016} can be seen as a dataspace. It extracts {\em metadata\/} about millions of datasets used in Google, including statistics but also information about who uses the data when. Unlike Goods, our focus is on integrating heterogeneous data in a graph. Goods is part of the {\em data lake} family, together with products from e.g., \href{https://www.ibm.com/analytics/us/en/data-management/data-lake/}{IBM} and \href{https://azure.microsoft.com/en-us/solutions/data-lake/}{Microsoft}. Our approach can be seen as consolidating a highly heterogeneous data lake in a single graph; we consider scaling up through distribution in the near future. 

{\em Information extraction} (IE) provides techniques to automatically extract structured information such as entities, relationships between entities, unique identifiers of entities, and attributes describing entities from unstructured sources \cite{DBLP:journals/ftdb/Sarawagi08}. 
The following two IE tasks are particularly relevant in our context.

\textit{Named-Entity Recognition} (NER, in short) is the task of identifying phrases that refer to real-world entities.
There exist a variety of NER tools. Firstly, we have web services such as Open Calais\footnote{\url{https://permid.org/onecalaisViewer}}, Watson NLU\footnote{\url{https://www.ibm.com/cloud/watson-natural-language-understanding}} etc.; 
such services limit the requests in a given time interval, making them less suitable for our needs. 
Libraries include Stanford NER~\cite{finkel2005incorporating} from Stanford CoreNLP~\cite{manning2014stanford}, Apache OpenNLP\footnote{\url{https://opennlp.apache.org/}}, SpaCy, 
KnowNER~\cite{seyler2018study} from the Ambiverse framework, and Flair~\cite{akbik2019flair}. These open-source libraries can be customized; they all support English, while Spacy and Flair also support French.
These NER tools use supervised techniques such as conditional random fields (Stanford NER) or neural architectures (Flair); text is represented either via hand-crafted features or using word embeddings. We adopted Flair since it is shown to perform well\footnote{\url{http://nlpprogress.com/english/named_entity_recognition.html}}.

\textit{Entity disambiguation} aims at assigning a unique identifier to entities in the text. In comparison to NER, there exist far fewer tools for performing disambiguation. From the tools mentioned above, the web-services that perform disambiguation are Gate Cloud and Dandelion, while from the third-party libraries, only Ambiverse and SpaCy have this functionality. SpaCy entity disambiguation is a supervised approach\footnote{https://spacy.io/universe/project/video-spacy-irl-entity-linking}, which represents a bottleneck, as there is no dataset annotated for French, to the best of our knowledge. The Ambiverse framework implements AIDA~\cite{hoffart2011robust}, an unsupervised technique for entity disambiguation. 


Overall, our work aims to facilitate the exploitation of reference data sources for computational journalism and journalistic fact-checking~\cite{cazalens:hal-01722666}; the problem is also related to truth discovery~\cite{DBLP:journals/jdiq/Berti-EquilleB16}.  Our graphs are a form of probabilistic graphs~\cite{DBLP:conf/pods/AmarilliMS17,DBLP:journals/tods/ManiuCS17}, which raise many hard algorithmic problems. The keyword search algorithm we use to exploit these graphs is described in~\cite{cordeiro:hal-02559688}, then \cite{query-paper}.

\vspace{2mm}
\noindent\textbf{Perspectives} In the near future, we will study a use case concerning conflicts of interest of medical experts, integrating different datasets from the medical literature (in XML and PDF). Following the intense evolution of the tool until its current shape, we plan to share it (as a .jar) with journalists so that they can use it directly.  

Other lines of research currently ongoing include: learning to rank query answers on our integrated graphs; simplifying the graphs through a form of abstraction; 
developing visual exploration interfaces (in collaboration with E.~Pietriga). The research into building and exploiting such graphs will continue as part of the SourcesSay ANR AI Chair project (2020-2024).

\vspace{2mm}
\noindent\textbf{Acknowledgements.} We thank Julien Leblay for his contribution
to earlier versions of this work~\cite{Chanial2018,cordeiro:hal-02559688}. We thank Xin Zhang for extracting from YAGO 4 the subset
used here. 

\renewcommand{\baselinestretch}{0.925}
{\small 
\bibliographystyle{abbrv}
\bibliography{the}}
\end{document}